\documentclass[prb,preprint]{revtex4-1}
\usepackage{graphicx}% Include figure files
\usepackage{dcolumn}% Align table columns on decimal point
\usepackage{bm}% bold math
\usepackage{amssymb,amsmath}

\begin{document}
\title{Universal Power-Law Strengthening in Metals?}
\author{P. M. Derlet}
\email{Peter.Derlet@psi.ch} 
\affiliation{Condensed Matter Theory Group, Paul Scherrer Institut, CH-5232 Villigen PSI, Switzerland}
\author{R. Maa{\ss}}
\affiliation{Institute for Materials Physics, University of G\"{o}ttingen, Friedrich-Hund-Platz 1, D-37077 G\"{o}ttingen, Germany}
\date{\today}

\begin{abstract}
The strength of most metals used in daily life scales with either an internal or external length scale. Empirically, this is characterized by power-laws persisting to six orders of magnitude in both strength and length scale. Attempts at understanding this scaling have generally been based on a specific mechanism. However the wide applicability of material type and microstructure to this phenomenon suggests a single mechanism is unlikely to capture the observed trend. Here we develop a model which gives quantitative insight into the scaling exponent using the known universal properties of a dislocation network and the leading order stress dependence of an underlying critical stress distribution. This approach justifies a value for the scaling exponent for virtually any experimental data set within the frameworks of both Hall-Petch strengthening and the ``small is stronger'' paradigm of small scale plasticity.
\end{abstract}


\maketitle

One of the scientifically most studied problems in the field of strength of crystalline materials is strengthening via grain size-reduction. This was experimentally demonstrated in the 1950s by both Hall~\cite{Hall1951} and Petch~\cite{Petch1954} for mild steel, ingot iron, spectrographic iron, as well as Zn. The stress, $\sigma$, at which strength is measured scales for all of these polycrystalline metals as $\sigma=\sigma_{0}+kd^{-n}$, where $\sigma_{0}$ is some base resistance of the constituting single crystal, $d$ is the grain size, and $k$ is commonly referred to as the Hall-Petch constant. The Hall-Petch exponent, $n$, is typically $\simeq0.5$. This empirical and technologically relevant scaling between strength and grain size appears simple, but has remained a fundamental challenge in metal physics. In fact, numerous mechanistic models  have been proposed (dislocation pile-up, work hardening, composite models, etc., see ref.~\cite{Lasalmonie1986} and references therein) to explain this scaling.

The persistence of the power-law scaling above a certain critical size $d$ is well reflected by the fact that not only very different microstructures (well annealed versus heavily cold rolled Ni, well annealed Fe, pearlitic steel, martensitic steel, tempered steel, etc.) obey Hall-Petch strengthening, but so do also fundamentally different parameters such as the yield strength, the lower yield point, the maximum flow strength, and the hardness or (see Petch~\cite{Petch1954}) the cleavage strength at $-198^{\circ}$C. Since the range of microstructures covers everything between low defect densities in large pure crystallites, and immensely complex hierarchical defect structures of advanced steels that include precipitates, carbides, various types of phases, grain and dislocation boundaries, lathe pockets and dislocation density gradients, it urges the question if any single mechanistic picture can be held responsible for this phenomenon?

A power-law strengthening with respect to a micro structural length scale is also seen in dynamic recrystallization~\cite{Derby1991} and recovery~\cite{Sherby1967}. Indeed, for the case of recrystallization, Derby~\cite{Derby1991} has demonstrated power law scaling for a wide range of materials including different grades of steels, Cu, Ni, Mg, Fe, FeS, and also the non-metals NaCl, NaNO$_{3}$, olivine and ice, with the exponent $n$ ranging between 0.5 and 0.8. This has also been shown for ultra fine grade metals with approximately $100<d<3000$ nm tested between $-196^{\circ}$C and $720^{\circ}$C~\cite{Pougis2013}. A third prominent example of power-law scaling is the ``smaller is stronger'' paradigm of micron- and nano-sized single crystals~\cite{Uchic2004,Uchic2009}. For this extrinsic size effect, where $d$ characterizes an external length scale, $n$ typically covers values between 0.2 and 0.7, and is for example found to depend on the initial dislocation density~\cite{El-Awady2013}.

\begin{figure}
\centerline{\includegraphics[width=.65\textwidth]{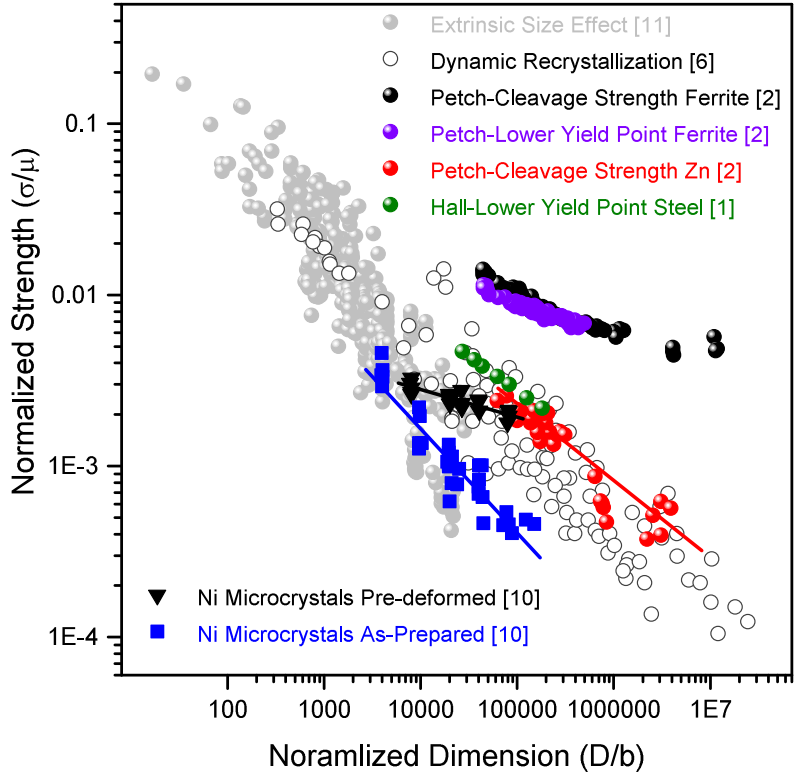}}
\caption{Log-Log plot of strength versus an internal or external length scale for a wide range of literature data for both small scale plasticity and grain size data, including the original data from both Hall and Petch. Following Derby~\cite{Derby1991}, the shear-strength versus length scale data is plotted in the respective units of an appropriate shear modulus and Burgers vector magnitude.}\label{FigOne}
\end{figure}

The above motivates fig.~\ref{FigOne}, which summarizes literature data for Hall-Petch strengthening~\cite{Hall1951,Petch1954}, dynamic recrystallization~\cite{Derby1991}, and size-affected strength (see for example ref.~\cite{Derlet2014} and references therein). Fig.~\ref{FigOne} demonstrates the remarkable fact that strength follows a similar power law with respect to both intrinsic (internal) and extrinsic (external) length scales for a vast range of materials and microstructure. 

In this research report we extend previous work~\cite{Derlet2014} rationalising the ``smaller is stronger'' paradigm as a general statistical sampling effect, to the much broader phenomenon of grain size strengthening and the Hall-Petch relation. The approach requires no specific mechanism (although none is discounted) and originates from only a knowledge that a dislocation network exhibits scale-free behaviour and that the extreme value statistics of a critical stress distribution is at play. By doing so, it is found that grain size strengthening, vis \'{a} vis the Hall-Petch mechanism, and the ``small is stronger'' paradigm are one and the same thing. This result also gives quantitative insight into the extent to which the scaling in strength is a truly universal phenomenon.

Ref.~\cite{Derlet2014} demonstrated that two quite different statistical effects contribute to the size effect in small scale plasticity, one occurring in stress and one in plastic strain. In what follows, only leading order trends are considered, an approach entirely compatible with the notion that logarithmic accuracy is sufficient for the emergence and identification of the size effect phenomenon in data-sets such as that shown in fig.~~\ref{FigOne}.

For the stress scaling, the internal dislocation network is characterized by a positive valued distribution, $P(\sigma)$, of critical stresses. Each such critical stress is the stress required for an irreversible rearrangement of the network and thus a plastic event. For a given elemental volume, $L^{3}$, there exists $M=\rho L^{3}$ such critical stresses ($\rho$ being the density of the available critical stresses). Sampling the distribution $M$ times gives a sequence of critical stresses, the smallest of which play the dominant role in initiating the transition to plastic flow. If $M$ is large then the statistics of the extreme controls these relevant critical stresses, whereas if $M$ is small then the statistics of the most probable becomes relevant. This rather general description naturally results in a shift to higher critical stresses when volume (and therefore $M$) decreases.
 
For sufficiently large $M$ ($M>100$), the apparatus of extreme value statistics defines the average $i$th critical stress, $\sigma_{i}$, of the ordered sequence via~\cite{Derlet2014} 
\begin{equation}
i=M\int_{0}^{\sigma_{i}}\mathrm{d}\sigma\,P[\sigma]. \label{Eqn1}
\end{equation}
The above is a generalization of the well known $i=1$ case of the average minimum value of a sampled ordered sequence of size $M$~\cite{Bouchaud1997}. For the small-stress regime, $P[\sigma]\sim\sigma^{\alpha}$, and eqn.~\ref{Eqn1} leads to $\sigma_{i}\sim (i/M)^{1/(1+\alpha)}\sim (i/L^{3})^{1/(1+\alpha)}$. Thus, as the volume reduces the stress scale increases. Apart from $\alpha$, this result is independent of the overall form of the positive valued distribution.

For strain scaling, the universal scaling properties of a dislocation network in a state of criticality is exploited~\cite{Zaiser2006}. In particular, like that of earth quakes, avalanche sizes and crackling noise (see for example refs.~\cite{Sethna2001,Mehta2006}), the distribution of strain magnitudes, $\delta\varepsilon$, associated with intermittent plasticity follows a power-law form with a non-algebraic scaling function (prefactor), $f[\cdot]$. Here $f[\cdot]$ depends on a length scale which in ref.~\cite{Derlet2014} characterized the sample volume. That is, $P[\delta\varepsilon]\sim f[\delta\varepsilon/\delta\varepsilon_{\mathrm{max}}(L)]\delta\varepsilon^{-\tau}$ where $\tau$ is a universal scaling exponent for intermittent plastic strain activity~\cite{Zaiser2006,Csikor2007} and $\delta\varepsilon_{\mathrm{max}}(L)$ varies inversely with $L$~\cite{Csikor2007,Zaiser2007}. Thus the plastic strain magnitude scale will be characterized by some function of $L$. Using a well-accepted representation of the scaling function~\cite{Derlet2014}, this characteristic scaling is found to be $\delta\varepsilon_{i}\sim L^{\tau-2}$, which gives the simple scaling of total plastic strain at the $i$th plastic event as $\varepsilon_{i}\sim i L^{\tau-2}$.

When put together, $\sigma_{i}\sim(i/L^{3})^{1/(1+\alpha)}\sim(\delta\varepsilon_{i}L^{2-\tau}/L^{3})^{1/(1+\alpha)}$, and the critical stress at a fixed plastic strain is found to scale as $L^{-(\tau+1)/(\alpha+1)}$ giving a size effect exponent of $n=(\tau+1)/(\alpha+1)$.

The above approach can be generalized to a polycrystalline material in a straight forward manner by considering an ensemble of grains, whose characteristic volume is defined as $L_{\mathrm{grain}}^{3}=d^{3}$. This also defines $M_{\mathrm{grain}}=\rho L_{\mathrm{grain}}^{3}$. The critical stresses available to the bulk material are described by a single effective distribution where the total number of critical stresses available is given by $M_{\mathrm{bulk}}=\rho L_{\mathrm{bulk}}^{3}$. Eqn.~\ref{Eqn1} then gives the $i$th average critical stress of the bulk polycrystalline system as $\sigma_{i}\sim (i/M_{\mathrm{bulk}})^{1/(1+\alpha)}$. 

To see how a single effective distribution of critical stresses may represent the extreme value statistics of critical stresses of a polycrystalline environment, eqn.~\ref{Eqn1} is generalized to
\begin{equation}
i=\sum_{n\in\mathrm{grains}}M_{n}\int_{0}^{\sigma_{i}}\mathrm{d}\sigma\,P_{n}[\sigma], \label{Eqn2}
\end{equation}
where the $n$th grain is characterized by its own critical stress distribution $P_{n}[\sigma]$ and $M_{n}$. $P_{n}[\sigma]$ is expected to depend on grain shape, the local grain network structure and also grain orientation with respect to the actual loading geometry defined by full the stress tensor $\sigma^{\mu\nu}$. The above equation may be written in the form of eqn.~\ref{Eqn1} with
\begin{equation}
M=\sum_{n\in\mathrm{grains}}M_{n}=M_{\mathrm{bulk}}
\end{equation}
and
\begin{equation}
P[\sigma]=\sum_{n\in\mathrm{grains}}\frac{M_{n}}{M}P_{n}[\sigma].
\end{equation}
Since $M_{n}$ is proportional to the grain volume, the effective critical stress distribution of the bulk polycrystalline material is given by a weighted volume average over the individual grain critical stress distributions. For an isotropic grain boundary network, such an average justifies the use of a scalar stress measure for $P[\cdot]$.

On the other hand, a plastic event within a grain, with an associated plastic strain $\delta\varepsilon_{\mathrm{Grain}}$, will admit a bulk plastic strain $\delta\varepsilon_{\mathrm{bulk}}\sim\left(L_{\mathrm{grain}}/L_{\mathrm{bulk}}\right)^{3}\delta\varepsilon_{\mathrm{grain}}=\left(M_{\mathrm{grain}}/M_{\mathrm{bulk}}\right)\delta\varepsilon_{\mathrm{grain}}$. This relation is a direct result of the work of Eshelby on irreversibly deformed inclusions~\cite{Eshelby1957}, and has also been verified in a polycrystalline environment via atomistic simulations of dislocation based nanocrystalline plasticity~\cite{Derlet2002}. Thus the plastic strain at the $i$th plastic event scales as $\varepsilon_{i}\sim i \left(M_{\mathrm{grain}}/M_{\mathrm{bulk}}\right)L_{\mathrm{grain}}^{\tau-2}$. It is immediately clear that the modifications to the stress and strain scaling cancel when combined, again giving a size effect exponent of $n=(\tau+1)/(\alpha+1)$.

The above derivation is formally applicable to the micro-plastic part of the deformation curve --- a regime of plastic deformation where significant structural change is minimal~\cite{Young1961,Vellaikal1969,Koppenaal1963}. We note that the original work by Hall and Petch examines quantities (lower yield point, cleavage strength) that were derived at the elastic-plastic transition. Indeed, Petch himself points out the absence of macroscopic plasticity upon recording the cleavage strength of Zn at $-198^{\circ}$C~\cite{Petch1954}. In the plastic flow regime, where dislocation network structure within the grains, grain boundary and grain boundary network structure can all evolve leading to the complex phenomenon of strain hardening and ductility, the distribution of critical stresses is expected to change. Moreover, it is well known that the non-universal aspects of the traditional Hall-Petch relation can depend on evolving material properties in the plastic flow regime~\cite{Hansen2004}. This is also the case for the ``smaller-is-stronger'' paradigm, where deviations from uniaxial boundary conditions lead to a strain hardening contribution rendering $n$ strain dependent~\cite{Maass2009}. Since the distribution of critical stresses will ultimately depend on micro-structure, the current work also suggests the exponent itself can evolve with structure.

\begin{figure}
\centerline{\includegraphics[width=.65\textwidth]{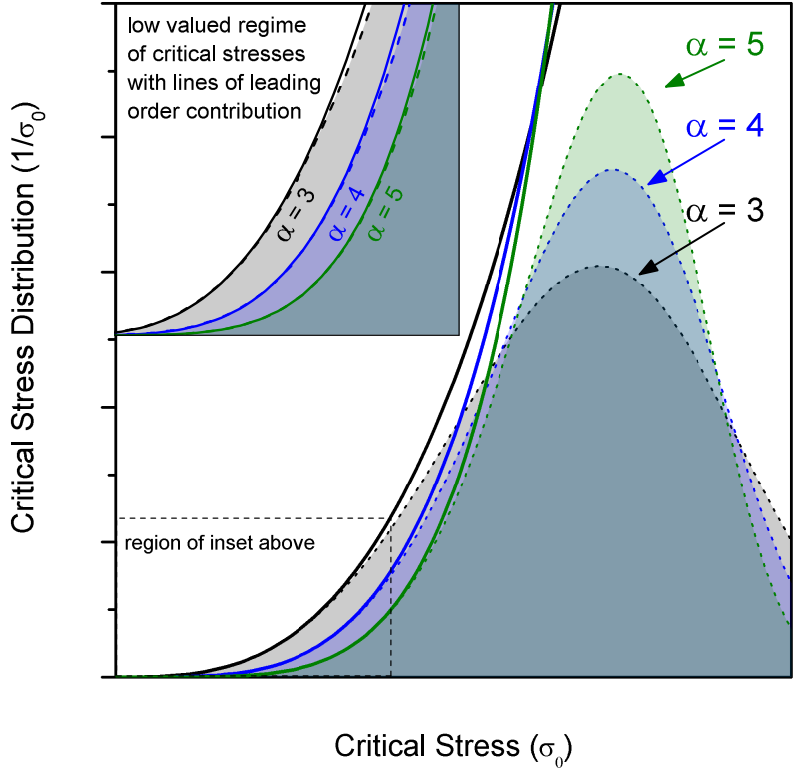}}
\caption{Plot of Weibull critical stress distributions (with shape parameter $\sigma_{0}$) for three different values of $\alpha$ able to reproduce the range of trends seen in fig.~\ref{FigOne} when the mean-field value of $\tau=3/2$ is used. The inset high-lights the low stress tail of each distribution and their leading order power law forms --- a regime of critical stresses which play the leading order role in the transition from elasticity to plastic flow.}\label{FigTwo}
\end{figure}

The assumption and use of a distribution of critical stresses and plastic strain increments embodies the premise that the physics of deformation within a grain differs little from that of a bulk single crystal containing a comparable dislocation network. While the application of this principle to micron-deformation is rather straight forward~\cite{Derlet2014}, its application to that of an isotropic polycrystal tacitly assumes that a statistically meaningful average over grain orientation and shape can be done (see appendix). The trends exhibited in fig.~\ref{FigOne} demonstrate that this must be the case --- an observation which is independent of any particular theory. For the case of micron-scale plasticity, some insight into the form of a critical stress distribution has been given by Isp\'{a}novity {\em et al} \cite{Ispanovity2013} using two and three dimensional dislocation dynamics simulations, and also experiment. They found that at different stages of their deformation curve the statistcs of their stress levels was close to a Weibull distribution with a Weibull exponent of $\alpha+1\simeq3.5-5.5$. Mean-field-theory gives the critical exponent associated with the plastic strain distribution as $\tau=3/2$ (see for example ref.~\cite{Zaiser2006,Mehta2006}). Numerical dislocation dynamics simulations have demonstrated this to be applicable to dislocation networks that admit either single or multiple slip plastic activity~\cite{Zaiser2006,Csikor2007,Derlet2013}. More recently Isp\'{a}novity {\em et al} \cite{Ispanovity2014} have demonstrated that dislocation dynamics simulations can also exhibit $\tau$ exponent values that are less than the mean-field value. Taking the mean-field prediction for the exponent $\tau=3/2$ with the usual Hall-Petch exponent of $n=1/2$ gives $\alpha=4$. The data of ref.~\cite{Ispanovity2013} is therefore quite compatible with the present theory. Fig.~\ref{FigTwo} plots the positive valued Weibull distribution of critical stresses with their leading order contributions at low stresses for $\alpha=3$, 4 and 5. Inspection of this figure demonstrates that for the regime of physically realistic distributions, the leading order contribution describes well the important low stress tail.

There must exist a lower grain-size limit to the applicability of the present work. Past experimental work on single crystals has shown that scale-free statistics remains operative down to sample volumes of several hundred nano-meters~\cite{Zaiser2008,Friedman2012}, and similarly that grain sizes down to 100 nm still follow the scaling summarized in fig.~\ref{FigOne}~\cite{Pougis2013}. However, at some low enough grain size, the surrounding grain boundary structure must begin to explicitly affect bulk plasticity~\cite{Gleiter1989}. In this regime of the nanocrystal, high-strain-rate atomistic simulations have shown that dislocations can nucleate at the grain boundary and contribute non-negligibly to plasticity (see ref.~\cite{Derlet2009} and references therein), and experimentally a break-down of the Hall-Petch effect is observed at the very smallest of grain sizes~\cite{Meyers2006}.

As with small-scale crystals and the ``smaller is stronger'' paradigm, for polycrystalline materials there must also exist an upper limit in $M$ (and therefore an upper grain diameter) at which the Hall-Petch effect becomes negligible. Indeed, when the grain size becomes sufficiently large, internal length scales within the dislocation network will naturally emerge and dominate the stress and strain statistics, thus decoupling the material's strength from the grain diameter. For polycrystalline materials the Hall-Petch effect becomes minimal at grain-sizes of $\sim100$ microns, where now $\sigma_{0}$ describes the strength ---  a limit which seems to be partly captured in the original data by Petch when displayed in fig.~\ref{FigOne}. In continuing the analogy to small-scale crystal experiments, it is noted that in the original work of Uchic {\em et al}~\cite{Uchic2004}, the authors found it surprising that the ``smaller is stronger'' paradigm remains operative at external length scales in the regime of tens of microns. This result now follows naturally from the current re-interpretation of the Hall-Petch data of fig.~\ref{FigOne}.

A question that still motivates continued efforts is why the exponent for confined plasticity is often higher than that seen in polycrystalline materials. The current body of small-scale crystal plasticity experiments though clearly evidences that when examining $n$ close to the break-away stress at low plastic strains, it rarely exceeds values above $\simeq0.6$, and decreases upon increasing the initial dislocation density. Whilst this is in good agreement with the here proposed model, instrumental effects in small-scale crystal experiments not present in bulk polycrystalline deformation, and the different boundary conditions (approximately open in the case of micro-pillars and approximately fixed in the case of polycrystals) are in fact expected to lead to some form of logarithmic correction to the exponent $\alpha$. This aspect must be investigated in future work.

By presenting a wide variety of experimental data, a very general power-law scaling emerges between material strength and a length scale which may be either intrinsic or extrinsic to the material. Although foremost an empirical power-law, its insensitivity to material type and microstructure nevertheless suggests a quite fundamental phenomenon is at play, which is not specific to any one particular mechanism. But is it strictly a universal phenomenon, as formally defined in ref.~\cite{Sethna2001}? While the developed expression for the strength exponent contains one universal exponent, $\tau$, the remaining exponent $\alpha$ is currently not considered to be universal (nor geometrical~\cite{Carpinteri2005}), ultimately depending albeit weakly on material type and microstructure. The present work therefore suggests universality partially underlies the phenomenon of size-strengthening, a status that could change upon the development of a quantitative theory of $\alpha$.

\begin{acknowledgments}
P.M.D. thanks P.D. Isp\'{a}novity for helpful discussion. R. M. thanks C.A. Volkert for institutional support.
\end{acknowledgments}

\end{document}